\documentclass[iop]{emulateapj}
\usepackage{amsmath}
\usepackage{epstopdf}
\usepackage{mathrsfs}
\usepackage{multirow}

\begin{document}
\title{Constraining the Frequency of Free-Floating Planets 
from a Synthesis of Microlensing, Radial Velocity, and Direct Imaging Survey Results}

\author{Christian Clanton\altaffilmark{1,2,3},
B. Scott Gaudi\altaffilmark{3}}
\altaffiltext{1}{NASA Postdoctoral Program Fellow}
\altaffiltext{2}{NASA Ames Research Center, Space Science \& Astrobiology Division, Moffett Field, CA 94035, USA}
\altaffiltext{3}{Department of Astronomy, The Ohio State University, 140 W. 18th Ave., Columbus, OH 43210, USA}
\email{christian.d.clanton@nasa.gov}

\begin{abstract}
A microlensing survey by \citet{Sumi2011} exhibits an overabundance of short-timescale 
events ($t_E\lesssim 2~$days) relative to that expected from known stellar populations 
and a smooth power-law extrapolation down to the brown dwarf regime.  This excess has 
been interpreted as a population of approximately Jupiter-mass  objects that outnumber 
main-sequence stars by nearly twofold; however the microlensing data alone cannot 
distinguish between events due to wide-separation ($a \gtrsim 10$~AU) and free-floating 
planets.  Assuming these short-timescale events are indeed due to planetary-mass 
objects, we aim to constrain the fraction of these events that can be explained by bound 
but wide-separation planets. We fit the observed timescale distribution with a lens mass 
function comprised of brown dwarfs, main-sequence stars, white dwarfs, neutron stars, 
and black holes, finding and thus corroborating the initial identification of an excess of 
short-timescale events. Including a population of bound planets with distributions 
of masses and separations that are consistent with the results from representative 
microlensing, radial velocity, and direct imaging surveys, we then determine what fraction 
of these bound planets are expected not to show signatures of the primary lens (host) 
star in their microlensing light curves, and thus what fraction of the short-timescale event 
excess can be explained by bound planets alone. We find that, given our model for the 
distribution of planet parameters, bound planets alone cannot explain the entire excess 
without violating the constraints from the surveys we consider, and thus some fraction of 
these events must be due to free-floating planets, if our model for bound planets holds. 
We estimate a median fraction of  short-timescale events due to free-floating planets to 
be $f = 0.67$ (0.23--0.85 at 95\% confidence) when assuming ``hot-start'' planet 
evolutionary models and $f = 0.58$ (0.14--0.83 at 95\% confidence) for ``cold-start'' 
models. Assuming a delta-function distribution of free-floating planets of mass 
$m_p=2~M_\mathrm{Jup}$ yields a number of free-floating planets per main sequence 
star of $N = 1.4$ (0.48--1.8 at 95\% confidence) in the ``hot-start'' case and $N = 1.2$ 
(0.29--1.8 at 95\% confidence) in the ``cold-start'' case.
\end{abstract}

\keywords{methods: statistical -- planets and satellites: general -- 
gravitational lensing: micro -- techniques: radial velocities -- 
techniques: high angular resolution -- stars: low-mass}

\section{Introduction}
\label{sec:sec1}

Deep optical and near-infrared photometric 
surveys to characterize the low-mass end of the substellar initial mass 
function (IMF) have identified populations of isolated, planetary-mass 
candidates in several nearby, young star-forming regions and clusters
\citep{Comeron1993,Nordh1996,Itoh1996,Tamura1998,Lucas2000,ZapateroOsorio2000,
ZapateroOsorio2002,McGovern2004,Kirkpatrick2006,Luhman2006,Bihain2009,
Burgess2009,Scholz2009,Scholz2012a,Scholz2012b,Weights2009,Marsh2010,Quanz2010,
Muzic2011,Muzic2012,Muzic2014,Muzic2015}. While a number of these 
photometrically-identified candidates have been met with some controversy in 
the literature concerning their youth (and thus low masses) or cluster 
membership \citep[see e.g.][]{Hillenbrand2000,Allers2007,Luhman2007a}, recent 
studies, in particular those by the Substellar Objects in Nearby Young Clusters 
(SONYC) group, have focused on obtaining confirmation spectra of such 
candidates and have verified several free-floating, planetary-mass objects 
with masses as low as a few Jupiter masses 
\citep{Scholz2009,Scholz2012a,Scholz2012b,Muzic2011,Muzic2012,Muzic2015}. 
\citet{Sumi2011} also present 
evidence for a large population of $\sim$Jupiter-mass objects that are 
either wide-separation ($a\gtrsim 10~$AU) or free-floating planets inferred from 
an excess of short events in the observed timescale 
distribution of a sample of microlensing events 
collected by the second phase of the Microlensing Observations in Astrophysics group 
\citep[MOA-II;][]{Sumi2003,Sako2008}. \citet{Wyrzykowski2015} report that data 
from the third phase of the Optical Gravitational Lensing 
Experiment \citep[OGLE-III;][]{Udalski2003} show a flattening in the slope of the 
observed event timescale distribution towards shorter timescales that is suggestive of 
a population of lenses similar to that reported by \citet{Sumi2011}, although this flattening  
is only marginally significant due to uncertainties resulting from small-number statistics 
and a low detection efficiency to such short-timescale events.

Comparing the occurrence rates of free-floating planets inferred by imaging 
surveys 
with those from microlensing is difficult, as the imaging surveys have 
sensitivities that cut off around 1--3~$M_\mathrm{Jup}$ (and depend on 
the exact evolutionary model adopted), while \citet{Sumi2011} 
found that these objects (regardless of their boundedness) most likely have 
masses near (and probably below) the sensitivity limit of the imaging surveys 
at $1.2^{+1.2}_{-0.7}~M_\mathrm{Jup}$. Nevertheless, in the SONYC survey of the 
young cluster NGC 1333, \citet{Scholz2012b} find that the occurrence rate of 
(photometrically-identified, spectroscopically confirmed) free-floating, 
planetary-mass objects relative to main-sequence stars is smaller than that 
inferred by the \citet{Sumi2011} microlensing study by a very large factor of 
some 20--50. \citet{Scholz2012b} argue that the star formation process extends 
into the planetary-mass regime, down to the planetary masses they are able to 
probe, and thus this large difference in inferred occurrence rates of 
free-floating planets must be due to a very large upturn 
in the mass function of compact objects 
below $\sim 3~M_\mathrm{Jup}$ that is 
perhaps indicative of a different formation channel (assuming that microlensing and 
direct imaging surveys are probing an analogous population of compact objects). 
Alternatively, one might argue that young open clusters may have a different 
mass function than the objects in the Galactic disk and bulge that give rise to microlensing 
events.

On the other hand, the photometric survey of $\rho$ Oph by \citet{Marsh2010} 
find a much larger number of isolated, planetary-mass objects per main-sequence 
star. After integrating their inferred mass function (shown in their Figure~8) in the 
planetary regime, the lowest two bins between 
$7\times 10^{-4} \lesssim M/M_{\odot} \lesssim 6\times10^{-3}$ (corresponding to 
roughly $0.7\lesssim M/M_\mathrm{Jup} \lesssim 6$), 
and in the stellar regime, the highest three bins between 
$0.08 \lesssim M/M_{\odot} \lesssim 1.0$, we divide these 
values to estimate the implied number of free-floating planets per main-sequence 
star of $\sim 30$. This number is over an order of magnitude larger than that 
of \citet{Sumi2011} and larger than the results of 
\citet{Scholz2012b} by an even greater factor. This seems to suggest that either the 
formation of free-floating 
planets is extremely sensitive to the local environment, the \citet{Marsh2010} sample 
is contaminated (with background stars or due to mis-estimates of the ages and/or 
masses of the candidate free-floating planets; see e.g. \citealt{Luhman2007a} and \citealt{Allers2007}) 
since they lack spectroscopic validation for many of their candidates, 
or some combination thereof.

Broadly, there are two formation channels for free-floating planets, 
but there are 
issues with the theory and observations behind each. The first, as 
\citet{Scholz2012b} claim, is that these objects form as an extension 
of the star formation process, however the lower mass fragmentation limit 
predicted by models of collapsing clouds is uncertain \citep[e.g.][]{Silk1977,Padoan1997,Adams1996} and may, in fact, 
be dependent on environment (e.g. \citealt{Bate2005}; also see 
\citealt{Luhman2007b} and \citealt{Bastian2010} for a discussion of the 
substellar IMF and its universality). Secondly, if free-floating planets initially form from material in 
circumstellar disks (either by disk fragmentation or core accretion), they 
must be subsequently ejected out of the system via dynamical processes such 
as planet-planet scattering, mass loss during post-main-sequence evolution, or 
ionization by interloping stars.

The ejection of a $\sim$~Jupiter-mass planet via planet-planet scattering requires a 
close encounter with another planet with a mass at least a Jupiter mass or above, as 
the least massive body in such an encounter is nearly always the one ejected 
\citep[see e.g. ][]{Ford2003,Raymond2008}.
Thus, if planet-planet scattering were the dominant channel for formation of 
the population of (presumably) free-floating, Jupiter-mass planets inferred by 
\citet{Sumi2011}, the frequency of Jupiter- and super-Jupiter-mass planets 
around low-mass stars must necessarily be high 
($\sim 50\%$; \citealt{Veras2012}), 
which is in significant disagreement with the predictions of core accretion 
theory \citep{Laughlin2004} as well as observational results from microlensing 
\citep{Gould2010,Cassan2012,Clanton2014b,Clanton2016}, radial velocity 
\citep{Bonfils2013,Montet2014}, and direct imaging 
\citep{Lafreniere2007,Bowler2015} surveys. Additionally, ejection due to 
mass loss during post-main-sequence evolution only works for planets with 
very wide orbital separations ($\sim$ several hundred AU) and requires 
(initial) host masses $\gtrsim 2~M_{\odot}$, and thus is not expected to 
produce free-floating planets at the required rate 
\citep{Veras2011,Mustill2014}.

Similarly, ionization by interloping stars requires initially wide 
planetary orbits and a dense stellar environment since the ionization time 
scales as $t_\mathrm{ion}\propto \nu^{-1}a^{-2}$, 
where $\nu$ is the local stellar 
number density and $a$ is the semimajor axis \citep[see][and references therein]{Antognini2016}. 
\citet{Antognini2016} demonstrate that even in the case of the most optimistic 
interaction cross sections, $t_\mathrm{ion}\sim 2~$Gyr, implying 
that $\sim 10\%$ of systems with planets on wide orbits would have been ionized 
in a cluster with an age of 200~Myr. In the field, these authors find 
$t_\mathrm{ion}\sim 4\times 10^{12}~$yr and therefore $\lesssim 1\%$ of 
wide-separation planetary systems would have been ionized in the lifetime of 
the Galaxy. Given current measurements of upper limits on the frequency of 
Jupiter- and super-Jupiter-mass planets with $a\gtrsim 10~$AU from direct 
imaging surveys of young FGK stars of $\lesssim 20-30\%$ 
\citep{Lafreniere2007,Biller2013} and young M stars of $\lesssim 16\%$ 
\citep{Bowler2015}, it does not seem likely that ionization (even in clusters) is 
able to produce the large numbers of free-floating planets inferred by 
\citet{Sumi2011}, although (to the best of our knowledge) a robust, 
quantitative analysis has yet to be performed.

Thus, while it may be possible to explain the formation of the smaller 
population of 
free-floating, planetary-mass objects observed by the SONYC group, the 
origin of the much larger 
population inferred by the \citet{Sumi2011} study remains elusive. One 
possible (and simple) 
solution could be that a majority of the planetary-mass objects 
needed to reproduce the over-abundance of short-timescale microlensing 
events seen in the MOA-II data \citep{Sumi2011} are not actually free-floating, 
but are gravitationally bound to host stars at wide enough orbital separations 
($a\gtrsim 10~$AU) that we do not expect to see signatures of the primaries 
(i.e. host stars) in a majority of their microlensing light curves and we do 
not expect them to be detected by direct imaging surveys (due to either lying outside the outer-working 
angles of such surveys, and/or having masses less than $\sim$~few Jupiter masses, below their detection limits).

In this study, we attempt to fit the observed timescale distribution with a 
standard lens mass function (hereafter LMF) comprised of brown dwarfs, main-sequence stars, 
white dwarfs, neutron stars, and black holes, along with a population of 
wide-separation, bound planets that is known to be 
consistent with the results of microlensing, radial velocity, and direct 
imaging surveys. In \citet{Clanton2016}, we demonstrated that there is a single 
planet population, modeled by a simple, joint power-law distribution 
function in planet mass and semimajor axis, that is simultaneously 
consistent with several 
representative surveys employing these three distinct detection techniques. 
Some fraction of such a planet population would produce detectable, 
short-timescale microlensing events that are well-fit by a single lens
model, similar in nature to the 10 observed events with $t_E<2~$days in the 
MOA-II data that \citet{Sumi2011} present.
We determine the expected timescale distribution for the combination of our 
adopted LMF and our planet population model and compare with the observed 
distribution to estimate the fraction of short-timescale events that are due to 
free-floating planets.

The remainder of the paper is organized as follows. We detail the properties of 
the \citet{Sumi2011} microlensing event sample and review their analysis to 
infer the existence of an abundant population of either wide-separation or 
free-floating planets in Section~\ref{sec:sec2}. We describe the different 
channels for distinguishing microlensing events due to free-floating planets 
from those due to bound planets in Section~\ref{sec:sec3}. We detail the 
methodologies we employ in this study in Section~\ref{sec:sec4} and present our 
results, together with discussion, in Section~\ref{sec:sec5}. Finally, we 
provide a summary of this work in Section~\ref{sec:sec6}.

\section{The Abundance of Wide-Separation or Free-Floating Planets Inferred by Microlensing}
\label{sec:sec2}
\citet{Sumi2011} select a sample of 474 well-characterized microlensing events 
from the 2006-2007 MOA-II data set. Here, well-characterized means that each 
light curve was determined to contain a genuine microlensing event that 
is distinguishable from intrinsically variable stars and other 
artifacts (e.g. cosmic rays, background supernovae). 
\citet{Sumi2011} require that each light curve have a single brightening 
episode consisting of more than three consecutive measurements (that are each 
$> 3\sigma$ above a constant baseline) and be ``well-fit'' by a theoretical 
microlensing model with a well-constrained (fractional error $\leq 0.5$) 
Einstein crossing time, $t_E$ (see 
Sections 2 and 3 and Table 2 of the Supplemental Materials of 
\citealt{Sumi2011} for a detailed description of their selection criteria and 
their particular definition of ``well-fit'').

Of these 474 microlensing events, 10 of them have timescales between 
$0.3\leq t_E/\mathrm{days}\leq 2$. 
For a lens mass $M_\mathrm{L}$, lens-source relative 
parallax $\pi_\mathrm{rel}$, and lens-source relative proper motion $\mu_\mathrm{rel}$, the 
Einstein crossing time scales as 
$t_E\propto (M_\mathrm{L}\pi_\mathrm{rel})^{1/2}/\mu_\mathrm{rel}$, which means 
that for typical values of $\pi_\mathrm{rel}$ and 
$\mu_\mathrm{rel}$\footnote{For a standard Galactic model 
\citep[e.g.][]{HanGould1995a,HanGould1995b,HanGould2003}, $\pi_\mathrm{rel}$ is 
expected to vary from $0.043~$mas to $0.21~$mas and $\mu_\mathrm{rel}$ is 
expected to vary from $4.2~\mathrm{mas~yr^{-1}}$ to $9.3~\mathrm{mas~yr^{-1}}$ 
for $68\%$ of events.}, microlensing events with 
timescales $t_E\lesssim 2~$days would indicate 
planetary-mass lenses.
Indeed, \citet{Sumi2011} fit the observed timescale distribution with an 
ensemble of simulated microlensing events 
appropriately weighted by their event rate 
as well as their detection efficiency as a function of $t_E$ 
(constructed by adopting a model of 
the Galaxy and a LMF over a mass range of 
$0.01\leq M_\mathrm{L}/M_\odot\leq 100$), and found an 
expected number of events with timescales $t_E<2~$days due to stellar, stellar 
remnant, and brown dwarf lenses to be either 1.5 or 2.5 (depending on their 
specific 
choice of form for the LMF). In either case, there is a clear 
overabundance 
of short-timescale microlensing events that is unexplained by such a model. 

\citet{Sumi2011} found that the fit to the overall 
timescale distribution is significantly improved when they included a 
population of planetary-mass objects as an extension to their canonical 
LMF (see their Figure~2).
\citet{Sumi2011} 
assumed that the population of planetary mass objects has a $\delta$-function 
mass 
distribution and found the value that most closely reproduces the observed 
timescale distribution to be $m_p=1.1^{+1.2}_{-0.6}~M_\mathrm{Jup}$. They also 
infer that the relative number of such objects to main-sequence stars 
($0.08\leq M_\star/M_\odot\leq 1.0$) is $1.9^{+1.3}_{-0.8}$ or 
$1.8^{+1.7}_{-0.8}$, again depending on the specific form of the mass function 
for the higher-mass lenses ($M_\mathrm{L}\geq 0.08~M_\odot$). 
\citet{Sumi2011} also tested a power-law mass function for the population 
of planetary-mass objects of the form 
$dN_\mathrm{pl}/d\log{m_p}=m_p^{1-\alpha_\mathrm{pl}}$ 
over the mass range $10^{-5}\leq m_p/M_\odot\leq 0.01$ (corresponding to 
$3~M_\oplus \lesssim m_p \lesssim 11~M_\mathrm{Jup}$) 
and found the slope that most closely 
reproduces the observed timescale distribution to be 
$\alpha_\mathrm{pl}=1.3^{+0.3}_{-0.4}$, from which they infer the relative 
number of planetary mass objects to main-sequence stars to be 
$5.5^{+18.1}_{-4.3}$. \citet{Sumi2011} note that while this power-law 
model has a maximum likelihood value that is 75\% smaller than that of their 
$\delta$-function planet mass model, it also has one fewer free parameter and 
is thus (formally) a slightly better fit. In the case of the $\delta$-function model, 
there are two additional free parameters, the mass and normalization, whereas 
in the case of the power-law model, the only additional free parameter is the slope 
(the normalization is included in the overall normalization of their LMF).

Although it is clear that a majority of the 10 events with $t_E<2~$days must be 
due to planetary-mass lenses (if they are indeed due to microlensing 
and standard models for the distributions of $\pi_\mathrm{rel}$ and 
$\mu_\mathrm{rel}$ are accurate), it is not certain whether these objects 
are gravitationally bound to a host star or if they are free-floating planets. 
\citet{Sumi2011} searched for signatures indicative of the presence of a host 
star in the light curves of the short-timescale events and found nothing 
(see Section~\ref{sec:sec3} for details on how to distinguish wide-separation 
planets from unbound planets), but 
were able to place limits on the projected 
separation (in units of the Einstein radius), $s$, of each planet 
from the host lens under the assumption that one exists (see their Table~1). 
These limits 
range between $2.4\leq s_\mathrm{min} \leq 15.0$, which roughly corresponds to 
semimajor axes between $6.7 \lesssim a_\mathrm{min} / \mathrm{AU} \lesssim 42$ 
assuming 
a typical primary lens, event parameters, and the median projection angle of a 
circular orbit, with a median value of $s_\mathrm{min}\simeq4.2$ 
($a_\mathrm{min}\simeq 12~$AU).
Here, the variables $s_\mathrm{min}$ and $a_\mathrm{min}$ 
represent the minimum values of the 
projected separation and corresponding semimajor axis (assuming a randomly-oriented, 
circular orbit), 
respectively, that would be plausible given the non-detection (at the $2\sigma$ level) 
of features 
in the microlensing light curves that would indicate the presence of a host star. 
We note that \citet{Sumi2011} did find three short-timescale events that clearly showed 
both binary lens caustic crossing features and very low-amplitude signals due to lensing 
by the primaries (see \citealt{Bennett2012} for an analysis of these three events, one of 
which was the first planetary microlensing event in which the host star was detected only 
through binary lensing effects, MOA-bin-1), but none of these 
passed all their selection criteria and made it into their final sample.

Since these microlensing data alone 
are insufficient to constrain the fraction of the population of planetary-mass 
lenses that are truly unbound, \citet{Sumi2011} consulted results from the 
Gemini Deep Planet Survey \citep[GDPS;][]{Lafreniere2007} that place upper 
limits on the frequency of wide-separation 
($10 \lesssim a / \mathrm{AU} \lesssim 500$) Jupiter- and super-Jupiter-mass 
planets. Using the information contained in Figure~10 of 
\citet{Lafreniere2007}, 
\citet{Sumi2011} estimated that $<40\%$ of the population of planetary-mass 
objects required to explain the overabundance of short-timescale microlensing 
events can be gravitationally bound to a host star at separations 
between $10-500~$AU, assuming any such planets have a uniform distribution 
of $\log{a}$.
 
However, we argue that the use of the full GDPS sample to constrain this fraction 
of bound planets 
is not correct (although we show in Section~\ref{sec:sec5} that our final 
result is actually consistent with the fraction estimated by \citealt{Sumi2011}). 
The stellar samples of the \citet{Sumi2011} survey and the GDPS are 
quite different, and thus the upper limits on planet frequency derived from the 
full GDPS sample are not necessarily representative of those for only the 
M stars.
This is an important point because microlensing samples are dominated by 
low-mass lens stars due to the fact that 
the rate of microlensing events depends explicitly on the mass function of 
lenses, which is weighted in favor of low-mass stars. On the other hand, the 
GPDS sample is comprised primarily of FGK stars, with a smaller number of M 
stars; of the full sample of 85 stars, just 16 are classified with M spectral 
types. While these stars are generally young, they are old enough that the 
lower-mass stars probably have spectral types that are not significantly 
different than that they will have when they fall on the main sequence. 
By comparing their observed $K$-band 
magnitudes to that predicted by stellar isochrones of M stars at similar ages, 
we argue that most, if not all, of the stars classified with M spectral types 
in the GDPS sample have a high likelihood of 
being an analogous population to the low-mass stars that produce 
the majority of microlensing events toward the Galactic bulge 
(see \citealt{Clanton2016} for discussion). This issue has been pointed out in 
\citet{Quanz2012}, who perform a more careful analysis using the GDPS 
constraints for just the M stars to estimate the upper limit on the fraction of 
the population of planetary-mass objects responsible for the observed 
short-timescale events that are bound to a host star, $f_\mathrm{max}$. 
These authors found a value of $f_\mathrm{max}=0.78$ (at 95\% confidence) 
if these planets have a typical mass of $1~M_\mathrm{Jup}$ and have 
separations equal to $a_\mathrm{min}$ that \citet{Sumi2011} calculate for each 
of the short-timescale events. If the planets are located at separations of 
$2a_\mathrm{min}$, then $f_\mathrm{max}=0.49$. Of course, the true planet 
population (assuming such a bound population exists) will have some 
distribution of separations, and will directly affect the value of 
$f_\mathrm{max}$. Another potential issue affecting both the \citet{Sumi2011} 
and \citet{Quanz2012} analyses is that the GDPS sensitivites (in terms of 
planet mass) they 
employ assume ``hot-start'' planet evolutionary models \citep{Baraffe2003}, 
which represent the most optimistic 
predictions for detecting planetary companions via direct imaging.

In this paper, we perform a thorough joint analysis of microlensing, 
radial velocity, and direct imaging constraints, selecting samples of stars 
similar to that probed by the \citet{Sumi2011} survey and considering 
both ``hot-'' and ``cold-start'' planet evolutionary models to determine
the expected timescale distribution of wide-separation, bound planets whose 
microlensing light curves reveal no evidence of the host stars they orbit. We 
will also do a more robust analysis than those of either \citet{Sumi2011} or 
\citet{Quanz2012} by including a distribution of planetary separations, 
including an outer cutoff semimajor axis for the population, to 
compute the fraction of the 
short-timescale events that are due to free-floating planets.

\section{Distinguishing Between Microlensing Events Due to Wide-Separation 
and Free-Floating Planets}
\label{sec:sec3}
In a microlensing event due to a wide-separation ($s\gg 1$, where $s$ is the 
projected separation in units of the Einstein radius) planet, evidence 
for boundedness can be obtained through three channels: 
1) observation of a relatively long-timescale (and likely low-magnification) 
bump due to the source trajectory 
passing near enough to the primary to produce a detectable magnification, 
2) observation of anomalies near the peak of the light curve due to the source 
passing near (or crossing) the planetary caustic \citep{Han2003}, and/or 
3) detecting blended light from the primary. 
The latter channel requires data of sufficient angular resolution to 
resolve out any unrelated stars, so that any additional flux above that of the 
source is due to the host lens (or a companion to the lens or source).
MOA-II data typically have seeing ranging between $1.9-3.5$~arcsec, with a median 
of $\sim 2.5$~arcsec \citep{Bond2001,Sumi2003}, and 
thus any detected blend flux could be (and is likely) due to unrelated stars, 
rather than the lens itself. The presence (or lack) of any blend flux in MOA-II 
data therefore provides no diagnostic power on the boundedness of 
planetary lenses.
Consequently, \citet{Sumi2011} were only able to look for evidence of a host 
lens in their 10 short-timescale events through the first two channels and so 
we do not need to consider the third channel in this study.

Detection of either a primary bump or anomalies due to the planetary caustic 
depends on the geometry of the event and the quality of 
the observations (e.g. total number of observations, cadence, photometric 
precision).
For a given set of observational (i.e. survey) parameters, the fractions 
of events due to wide-separation planets for which the presence of a primary is 
expected to be detected by these two different channels scales as 
$\sim s^{-1}$ and $\sim s^{-2}$, respectively. In this section, we provide 
brief descriptions of these two channels and how we implement them in this 
study, but for a more in-depth look at distinguishing events due to 
wide-separation and free-floating planets, see \citet{Han2005} and references 
therein.


\subsection{Low-Magnification Primary Bumps}
\label{subsec:sec3.1}
\begin{figure}[!h]
	\epsscale{1.15}
	\plotone{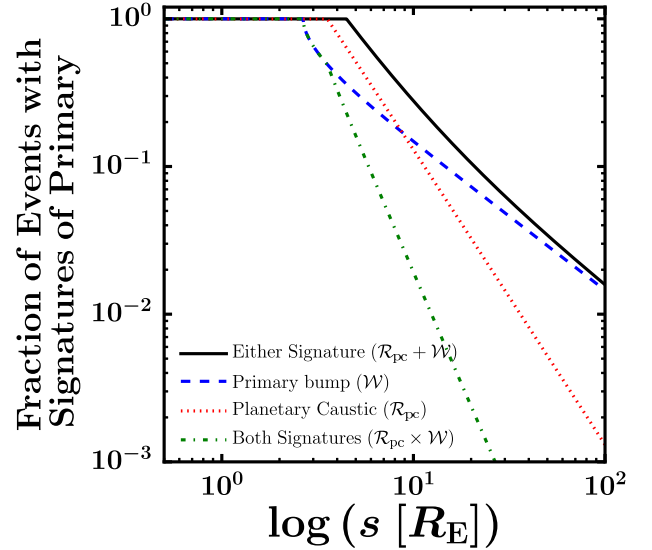}
	\caption{Fraction of planetary microlensing events for which signatures of 
	primary are expected to be detectable , i.e., the fraction of planetary 
	events we can distinguish as being due to bound planets rather than 
	free-floating planets (assuming $M_\mathrm{L}=0.3~M_{\odot}$, 
	$f_\mathrm{obs}=50~\mathrm{day^{-1}}$, 
	$\sigma_\mathrm{p}=0.05$, and 
	$\mathcal{Q}_\mathrm{th}=80$). We examine two channels for detecting 
	signatures of the primary: 1) relatively long-timescale, low-magnification 
	primary bump, and 2) anomalies near the peak of the light curve due to the 
	source passing near (or crossing) the planetary caustic. The relative 
	probabilities of these channels fall off with increasing projected separation 
	as $\mathcal{W}(s)\sim s^{-1}$ and $\mathcal{R}_\mathrm{pc}\sim s^{-2}$, 
	respectively. See text for a more detailed description.
		\label{fig:rpc_ws_plot}}
\end{figure}

The source trajectory in some fraction of planetary microlensing events, 
$\mathcal{W}(s)$, will be 
such that the impact parameter to the \textit{primary}, $u_{\star}$, is small 
enough that 
it will produce a detectable magnification. For a given observational cadence 
and signal-to-noise ratio (S/N; $\mathcal{Q}$) threshold, this fraction depends 
solely on the geometry of the lens system and has the form
\begin{equation}
  \mathcal{W}(s) = 
  \begin{cases}
	  \displaystyle 1 \; , & s\leq u_\mathrm{\star, th} \; , \vspace{0.2cm} \\
	  \displaystyle\frac{2}{\pi}\sin^{-1}\left(\frac{u_\mathrm{\star, th}}{\hat s}\right) \; , & s > u_\mathrm{\star, th} \; ,
  \end{cases}
  \label{eqn:w_s}
\end{equation}
where $\hat s \equiv s - 1/s$ is the projected separation of the center of 
the planetary caustic from the host star and $u_\mathrm{\star, th}$ is the maximum source 
impact parameter to the primary in units of the primary Einstein radius, 
$\theta_\mathrm{E}$, such that the primary bump is just detectable. 
The form for the maximum impact parameter we adopt is given by \citet{Han2005} 
as
\begin{align}
  u_\mathrm{\star, th} = & {} \; \displaystyle 2.2\left(\frac{M_\mathrm{L}}{0.3~M_{\odot}}\right)^{1/14} \left(\frac{f_\mathrm{obs}}{50~\mathrm{day^{-1}}}\right)^{1/7}\times \nonumber \\
    & {} \; \left(\frac{\sigma_\mathrm{p}}{0.05}\right)^{-2/7}\left(\frac{\mathcal{Q}_\mathrm{th}}{80}\right)^{-2/7}\; , 
  \label{eqn:u_star_th}
\end{align}
where $M_\mathrm{L}$ is the primary lens mass, $f_\mathrm{obs}$ is the 
frequency of observations, $\sigma_\mathrm{p}$ is 
the fractional photometric precision of each observation, and 
$\mathcal{Q}_\mathrm{th}$ is the S/N threshold for detection. 
The values to which we have scaled this relation are set by the selection 
criteria of the \citet{Sumi2011} study and are typical for MOA-II data 
(T. Sumi, private communication). 
For the given survey parameters ($f_\mathrm{obs}$, $\sigma_\mathrm{p}$, and 
$\mathcal{Q}_\mathrm{th}$) and at fixed primary lens mass ($M_\mathrm{L}$), 
we find that 
$\mathcal{W}(s) = 1$ out to projected separations $s\lesssim 2.6$. In the limit 
of large projected separations, $s\gg 1$, the probability of detecting the 
primary through this channel falls off as $s^{-1}$. 
We note that $u_\mathrm{\star, th}$ is only weakly dependent on 
$f_\mathrm{obs}$, $M_\mathrm{L}$, and $\sigma_\mathrm{p}$ and thus argue that 
our approximation that these parameters are the same for all events is reasonable.
Figure~\ref{fig:rpc_ws_plot} shows a plot of $\mathcal{W}(s)$ and illustrates 
the that, for parameters typical of the MOA-II survey and the selection criteria 
set by \citet{Sumi2011}, this is not the primary channel for detecting 
signatures of the host lens except at separations beyond $s\gtrsim 10$ 
($a\gtrsim 28~$AU).

\subsection{Anomalies Due to the Planetary Caustic}
\label{subsec:sec3.2}
The rate of planetary events where signatures of the primary due to anomalies 
near the peak of the light curve arising from the source passing near (or 
crossing) the planetary caustic relative to the total rate of planetary 
microlensing events is  
\begin{equation}
	\mathcal{R}_\mathrm{pc}(q,s) = \displaystyle \frac{u_\mathrm{pl, max}(s)}{\bar u_\mathrm{S11}} \; ,
	\label{eqn:R_pc_first}
\end{equation}
where $u_\mathrm{pl, max}(s)$ is the maximum required impact parameter 
for signatures of the planetary caustic to be just detectable as 
a function of the projected separation, $s$, and 
$\bar u_\mathrm{S11}=0.153$ is the 
median impact parameter measured by \citet{Sumi2011} for the 10 short-timescale 
($0.3\leq t_\mathrm{E} / \mathrm{day} \leq 3$) events in their sample. If 
the MOA-II survey were uniformly sensitive to events with respect to impact 
parameter, then we would have chosen to normalize $\mathcal{R}_\mathrm{pc}$ by 
$u_0=1$, the maximum impact parameter allowed by the criteria set by 
\citet{Sumi2011}, with which they selected their sample (see Section 2 of the 
supplemental materials of \citealt{Sumi2011}). In reality, there is a bias 
towards smaller impact parameters (since the total magnification, $A$, 
depends on the 
lens-source projected separation, $u$, as $A(u)=[(u^2+2)/(u\sqrt{u^2+4})]$) 
and we therefore attempt to account for this 
by normalizing $\mathcal{R}_\mathrm{pc}$ by $\bar u_\mathrm{S11}$.

We assume that in order for signatures of the planetary caustic to be 
detectable in the light curve that 
$u_\mathrm{pl, max}\sim \theta_\mathrm{c}$, 
where $\theta_\mathrm{c}$ is the angular radius of the planetary caustic, 
which we assume to be circular in shape with a size given by the height of the caustic in the 
direction perpendicular to the star-planet axis. 
Adapting equation~9 of \citet{Han2006} to be consistent with our adopted geometry, 
we find the following expression for $\theta_c$ (which has units of the primary Einstein radius) 
\begin{equation}
\theta_\mathrm{c} = \displaystyle \frac{2}{s\sqrt{s^2+1}}\; .
\label{eqn:pc_radius}
\end{equation}
There is a projected separation, $s_\mathrm{c}$, at which 
$\theta_\mathrm{c}>\bar u_\mathrm{S11}$ and interior to which 
$\mathcal{R}_\mathrm{pc}$, as defined by equation~(\ref{eqn:R_pc_first}), 
becomes greater than unity. This works out to be 
$s_\mathrm{c} \simeq 3.5$, which roughly corresponds to a projected separation 
in physical units of $r_\mathrm{\perp, c}\approx 10$~AU for typical event 
parameters and a semimajor axis of $a_\mathrm{c}\approx 12~$AU for the median 
projection angle of a circular orbit. Thus, to ensure that 
$\mathcal{R}_\mathrm{pc}\leq 1$, we adopt the definition 
\begin{equation}
	u_\mathrm{pl, max}(s) = 
	\begin{cases}
		\bar u_\mathrm{S11} \; , & s \leq s_\mathrm{c} \; , \\
		\theta_\mathrm{c} \; , & s > s_\mathrm{c} \; ,
	\end{cases}
	\label{eqn:u_pl_max_s}
\end{equation}
and equation~(\ref{eqn:R_pc_first}) takes the form 
\begin{equation}
	\mathcal{R}_\mathrm{pc}(q,s) = 
	\begin{cases}
		1 \; , & s \leq s_\mathrm{c} \; , \\
		\displaystyle \frac{\theta_\mathrm{c}}{\bar u_\mathrm{S11}} \; , & s > s_\mathrm{c} \; .
	\end{cases}
	\label{eqn:R_pc}
\end{equation}
In doing so, we are effectively assuming that if a planetary event 
with $s\leq s_\mathrm{c}$ is detected, 
anomalies due to the planetary caustic will always be detected.
For our purposes, this is not a problem, since we are only concerned 
with computing the fraction of events for which we expect to see evidence of a 
primary (regardless of the exact channel). Figure~\ref{fig:rpc_ws_plot} 
illustrates our expectation that for $s\lesssim 10$, perturbations due to the 
planetary caustic are the primary channel for revealing the presence of a 
host star.

\begin{table*}[!ht]
	\caption{\label{tab:tab1} Median values and 68\% uncertainties inferred by 
	\citet{Clanton2016} for  
	the parameters of a population of planets that is consistent with results 
	from the microlensing surveys of \citet{Gould2010} and \citet{Sumi2010}, the 
	Gemini Deep Planet Survey \citep{Lafreniere2007} and Planets Around Low Mass 
	Stars \citep{Bowler2015} direct imaging surveys, and the CPS TRENDS 
	\citep{Montet2014} RV survey.}
	\centering
	\begin{tabular}{c||c|c|c|c}
		\hline \hline
		Planet Evolutionary  &  \multicolumn{4}{c}{Median Values and 68\% Uncertainties} \\
		Model &  $\alpha$  & $\beta$  & $\mathcal{A}~[{\rm dex^{-2}}]$  & $a_{\rm out}~[{\rm AU}]$ \\
		\hline
		\begin{tabular}{@{}c@{}} ``Hot-Start'' \\ (\citealt{Baraffe2003})\end{tabular} & $-0.86^{+0.21}_{-0.19}$& $1.1^{+1.9}_{-1.4}$ & $0.21^{+0.20}_{-0.15}$ & $10^{+26}_{-4.7}$ \\
		\hline
		\begin{tabular}{@{}c@{}} ``Cold-Start'' \\ (\citealt{Fortney2008})\end{tabular} & $-0.85^{+0.21}_{-0.19}$& $1.1^{+1.9}_{-1.3}$ & $0.21^{+0.20}_{-0.15}$ & $12^{+50}_{-6.2}$ \\
		\hline\hline
	\end{tabular}
\end{table*}

\section{Methodology}
\label{sec:sec4}
In \citet{Clanton2016}, we performed a joint analysis of results from five 
different surveys for exoplanets employing three independent discovery 
techniques: microlensing \citep{Gould2010,Sumi2010}, 
radial velocity (specifically, the long-term trends; \citealt{Montet2014}), 
and direct imaging \citep{Lafreniere2007,Bowler2015}. We found that the results 
of all these surveys can be simultaneously explained by a single population of 
planets described by a with a simple, joint power-law distribution in mass and 
semimajor axis given by 
\begin{equation}
  \frac{d^2N_\mathrm{pl}}{d\log{m_p}~d\log{a}} = \mathcal{A}\left(\frac{m_p}{M_\mathrm{Sat}}\right)^\alpha\left(\frac{a}{2.5~\mathrm{AU}}\right)^\beta \; .
  \label{eqn:planet_dist_function}
\end{equation}
This model has just four free parameters, 
$\left\{\alpha, \beta, \mathcal{A}, a_\mathrm{out}\right\}$, where $a_\mathrm{out}$ is 
the outer cutoff radius of the semimajor axis distribution. The median values 
and 68\% confidence intervals we infer for these parameters are summarized in 
Table~\ref{tab:tab1}. Note that the quoted uncertainties, particularly those on 
$\beta$ and $a_\mathrm{out}$, are correlated (see Figures~25--27 in 
\citealt{Clanton2016}). In this paper, we employ this population of bound planets 
that is known to be consistent with microlensing, RV (long-term trend detections), 
and direct imaging surveys to explain (at least a significant fraction) of the 
overabundance of short-timescale microlensing events observed in the MOA-II data, 
and thus derive constraints on the frequency of truly 
free-floating planets in the Galaxy.

\begin{figure*}[!t]
	\epsscale{1.0}
	\plotone{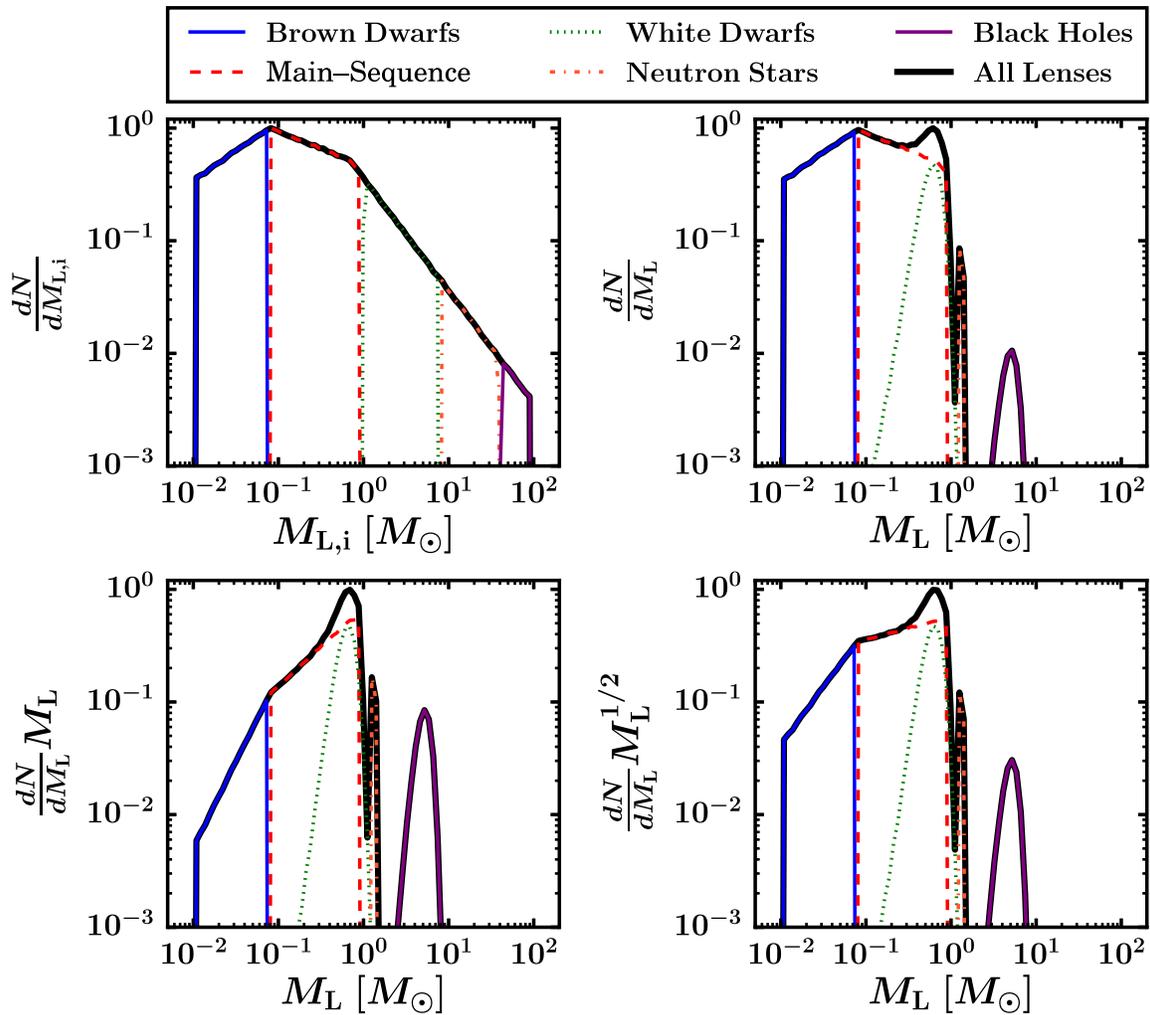}
	\caption{The lens mass function we adopt in this study (identical to ``Model 1'' 
	     of \citealt{Sumi2011}), consisting of 
		populations of brown dwarfs, main-sequence stars, white dwarfs, 
		neutron stars, and black holes, each of which is described by a 
		power-law distribution in their initial mass. The top left panel plots 
		the initial lens mass function, and the top right panel plots the final 
		lens mass function. The bottom left and bottom right panels show the 
		final mass function, weighted by lens mass and event rate 
		($\Gamma\propto M_L^{1/2}$ along a given sight line and at fixed $D_l$, 
		$D_s$, and $\boldsymbol{\mu}_\mathrm{rel}$), respectively.
		\label{fig:adopted_lmf}}
\end{figure*}

We first sample the posterior distributions (including covariances) 
derived in \citet{Clanton2016} to 
obtain parameters (i.e. $\alpha$, $\beta$, $\mathcal{A}$, and $a_\mathrm{out}$) 
for a random population of (bound) planets, and draw an ensemble of 
planets from 
the resultant distribution function. We then generate a corresponding 
set of simulated 
microlensing events, precisely following the procedure we outline in
\citet{Clanton2014a,Clanton2014b,Clanton2016}, but with a slightly altered 
LMF. In this paper, we adopt ``Model~1'' exactly as it is presented in 
\citet{Sumi2011}, 
which includes populations of brown dwarfs, main-sequence stars, white dwarfs, 
neutron stars, and black holes that are described by power-law distributions in their 
initial mass. 
We fix the slope of the LMF in the brown dwarf regime to be the median value 
reported by \citet{Sumi2011}, as the inferred value for this slope is not 
significantly different when the fitting the full timescale distribution versus 
fitting the timescale distribution for $t_E>2~$days (which we verified with our 
own, completely independent, fitting procedures). We display a plot of the 
initial lens mass distribution 
(relevant for the remnant populations) in Figure~\ref{fig:adopted_lmf}, 
along with plots of the final LMF weighted 
by number, mass, and contribution to the microlensing event rate along a given 
line of sight.
The relative 
numbers of brown dwarfs, main-sequence stars, white dwarfs, neutron 
stars, and black holes by number, mass, and event rate are 
(38:52:9.5:1.1:0.16), (5.9:63:22:5.7:3.2), and 
(17:63:17:2.9:0.84), respectively
(consistent with 
\citealt{Gould2000}). We find, as did \citet{Sumi2011}, that the numbers of 
brown dwarfs, white dwarfs, neutron stars, and black holes relative to 
main-sequence stars are (73:18:2.1:0.31).

\begin{figure}[!t]
	\epsscale{1.15}
	\plotone{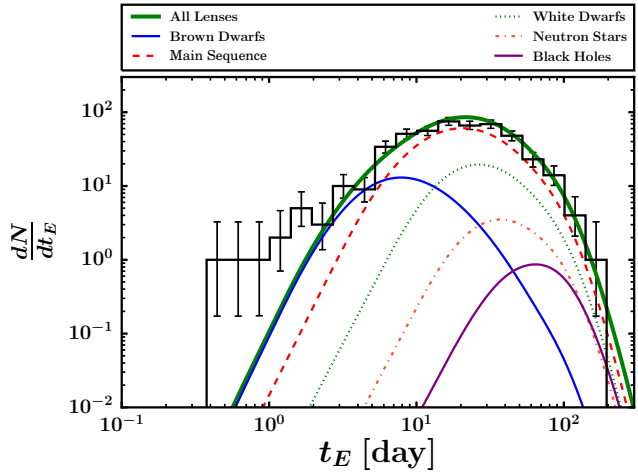}
	\caption{Predicted timescale distribution for populations of brown dwarfs, 
	    main-sequence stars, and stellar remnants (colored lines) and the observed 
	    timescale distribution reported by \citet{Sumi2011} (black histogram). The 
	    predicted timescale distribution has been subjected to the measured 
	    detection efficiency 
	    of the \citet{Sumi2011} survey and normalized to the total number of observed 
	    microlensing events. The number of short-timescale microlensing events 
	    ($t_E\leq 2~$days) predicted by our adopted LMF is 1.1, compared 
	    to the observed number of 10, demonstrating 
	    a clear overabundance of such short-timescale events in the observed sample.
		\label{fig:lmf_predicted_te_dist}}
\end{figure}

Figure~\ref{fig:lmf_predicted_te_dist} shows a plot of the predicted distribution 
of timescales for the brown dwarf, 
main-sequence, and remnant lenses in our simulated sample against the 
observed distribution. Our predicted distribution has been corrected for the 
detection efficiency determined by \citet{Sumi2011} (shown in Figure~S2 
of their supplementary materials) and normalized such that the total number of 
simulated microlensing events matches that of the observed sample. 
Note that this is not a fit to the observed 
distribution, but rather, it is a prediction 
based on a fit performed by \citet{Sumi2011} that we use to fix the slope of 
the LMF in the brown dwarf regime. 
By eye, this appears to be a good match for events with $t_E>2~$days 
(providing a degree of confidence in our adopted Galactic model and LMF), 
but the overabundance of shorter-timescale 
events in the observed distribution is clear. We will attempt to explain these 
short-timescale events with bound planetary companions for which we do not 
expect to see evidence of a primary in the microlensing light curves.

Having generated a population of planets with corresponding microlensing 
events as described above, we then determine the probability that the 
primary (i.e. host star) 
would \textit{not} be detected in each event given the survey parameters of MOA-II, 
$P_{\star}^{'} = [1 - \mathcal{W}(s)] \times [1 - \mathcal{R}_\mathrm{pc}(q,s)]$, where 
$\mathcal{W}(s)$ is the fraction of events where a low-magnification primary 
bump is expected to be detectable and $\mathcal{R}_\mathrm{pc}(q,s)$ 
is the fraction of 
events where perturbations in the light curve due to the planetary caustic are 
expected (see Section~\ref{sec:sec3} for the formal definitions and a 
discussion of these quantities). We then construct the predicted timescale 
distribution for the combination of our adopted LMF and the associated 
population of bound 
planets that \textit{appear} to be free-floating, again taking care to correct 
for the detection efficiency of MOA-II to events as a function of $t_E$. This 
predicted timescale distribution serves as our likelihood function (for which 
there is no analytic form). We calculate the likelihood of a given planet 
population by applying this numerically-generated likelihood function to the 
individual measurements of $t_E$ for each of the 474 events comprising the 
observed distribution presented in \citet{Sumi2011}. These data are published 
in Table~4 of \citet{Sumi2013}.
We repeat this procedure for all planet populations 
\citet{Clanton2016} found to be consistent with radial velocity, microlensing, 
and direct imaging surveys. This allows us to place constraints on the fraction 
of short-timescale ($t_E < 2~$days) microlensing events due to free-floating planets. 
In order to determine an actual number of such planets (e.g. relative to main-sequence 
stars), 
we must adopt an ad hoc form for the mass function of free-floating planets. 
Therefore, our estimate of the number of free-floating planets per star is less robust (i.e. 
more model dependent) than our estimate of the fraction of short-timescale events due to 
free-floating planets. 
We present and discuss our results and main sources of uncertainty in the following section.

\section{Results and Discussion}
\label{sec:sec5}
\begin{figure}[!t]
	\epsscale{1.15}
	\plotone{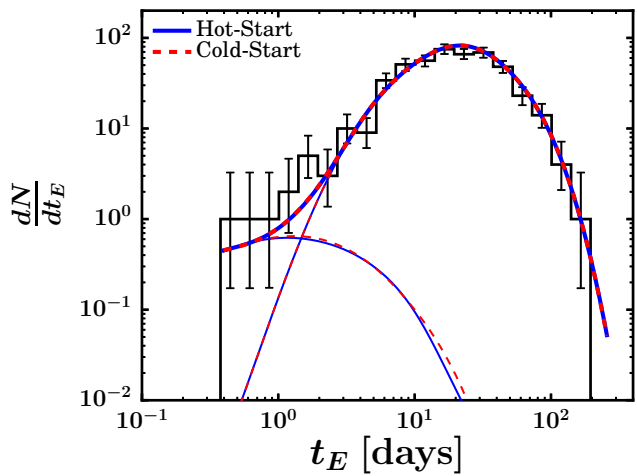}
	\caption{Maximum likelihood fits to the observed timescale distribution 
	              \citep[black histogram;][]{Sumi2011} for our canonical LMF and 
	              a population of bound, 
	              wide-separation planets that is consistent with results from radial 
	              velocity, microlensing, and direct imaging surveys \citep{Clanton2016}, 
	              assuming either  
	               ``hot-start'' \citep[blue lines;][]{Baraffe2003} or ``cold-start'' 
	               \citep[red lines;][]{Fortney2008} planet evolutionary models. The thick 
	               lines show the expected timescale distribution from all lenses, while 
	               the thin lines show the expected contributions from planets (the curves 
	               peaking at 
	               shorter timescales) and brown dwarfs, main-sequence stars, and 
	               remnants (the curves peaking at longer timescales). For these  
	               maximum likelihood 
	               fits, wide-separation, bound planets account for roughly 2.9 of the 10 
	               observed 
	               short-timescale ($t_E < 2~$days) events in both the ``hot-start'' and 
	               ``cold-start'' cases, and brown dwarfs account for about 
	               one event.
		\label{fig:max_lhood_te_dist_fit}}
\end{figure}

Figure~\ref{fig:max_lhood_te_dist_fit} shows the best-fit (i.e. 
maximum likelihood)
expected timescale distribution for the combination of our canonical LMF 
described in the previous section with populations of wide-separation, bound 
planets found by \citet{Clanton2016} to be consistent 
with results from radial velocity, microlensing, and direct imaging surveys for either 
``hot-start'' \citep{Baraffe2003} or ``cold-start'' \citep{Fortney2008} planet 
evolutionary models.
Given that the parameters of these planet populations (i.e. the slopes of the mass 
function, $\alpha$ and semimajor axis function, $\beta$, normalizations, 
$\mathcal{A}$, and outer cutoff radii, $a_\mathrm{out}$) for the ``hot-start'' and 
``cold-start'' models are not too different \citep[see Section~5.2 of][]{Clanton2016}, 
it is not surprising that the fits for these different models shown in 
Figure~\ref{fig:max_lhood_te_dist_fit} are so similar. The parameter values for the 
best-fit ``hot-start'' planet population are 
$\alpha=-0.85$, $\beta=0.091$, $\mathcal{A}=0.26~\mathrm{dex^{-2}}$, 
and $a_\mathrm{out}=740~$AU, and those for the best-fit ``cold-start'' population 
are similar. The value of $a_\mathrm{out}$ for this best fit is quite large due to the 
fact that the number of planets for which we do not expect to see signatures of a 
primary lens (host star) in the microlensing light curves (which are needed to 
explain the overabundance of short-timescale events) increases with this outer 
cutoff radius. For large $a_\mathrm{out}$, planets are allowed to be in very 
wide-separation orbits which lead to smaller planetary caustic sizes (and thus 
small rates of planetary caustic events, since at fixed $q$, 
$R_\mathrm{pc} \sim \theta_c \propto s^{-2}$ for $s\gg 1$) and which have low 
probability for source trajectories that pass near the primary 
($\propto s^{-1}$ for $s\gg 1$). However, in order for a planet population with a 
large value of $a_\mathrm{out}$ to be consistent with the non-detections from 
direct imaging surveys (i.e. \citealt{Lafreniere2007} and \citealt{Bowler2015}), the 
slope of the semimajor axis distribution function must be shallow, and indeed, the 
best-fit population has $\beta$ near zero (corresponding to \"Opik's law; 
\citealt{Opik1924}).

\begin{figure}
	\epsscale{1.15}
	\plotone{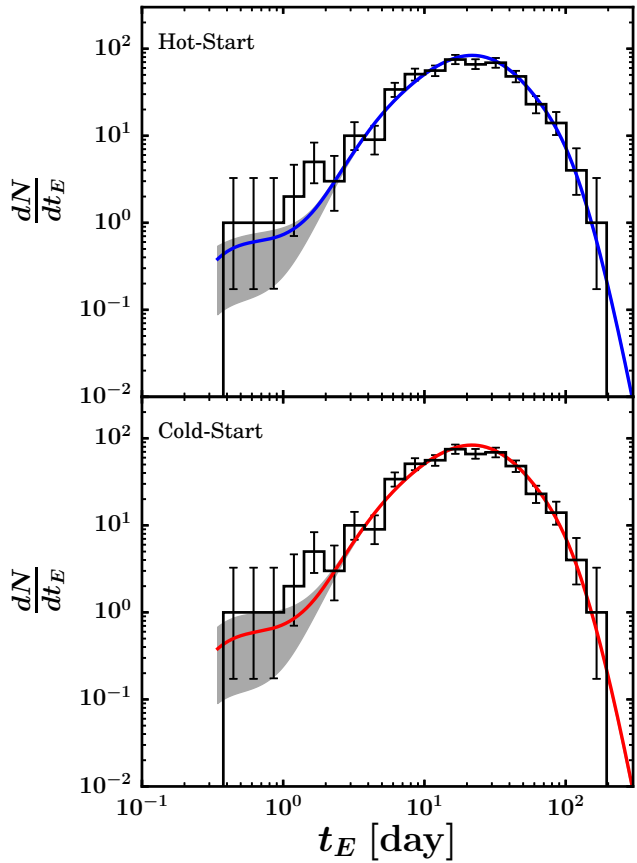}
	\caption{Maximum likelihood and 
	              68\% confidence interval fits to 
	              the observed timescale distribution 
	              \citep[black histograms;][]{Sumi2011} for our canonical LMF and 
	              a population of bound, 
	              wide-separation planets that is consistent with results from radial 
	              velocity, microlensing, and direct imaging surveys \citep{Clanton2016}, 
	              assuming either  
	               ``hot-start'' \citep[top panel;][]{Baraffe2003} or ``cold-start'' 
	               \citep[bottom panel;][]{Fortney2008} planet evolutionary models.
		\label{fig:te_dist_fit_range}}
\end{figure}

In Figure~\ref{fig:te_dist_fit_range}, we display the best-fit to the observed 
timescale distribution along with the range of fits in the 68\% confidence interval. It is 
clear from this figure that while we can explain some fraction of the short-timescale 
events with bound, wide-separation planets, an overabundance remains (particularly 
at timescales between 1--2~days). This suggests that either our assumed planet 
population model is incorrect in regions of parameter space where we currently have no 
observational constraints ($m_p \lesssim M_\mathrm{Jup}$ at separations 
$a\gtrsim 10~$AU), or free-floating planets are responsible for the remaining 
short-timescale events. We have no way of testing the former, but for the latter case 
we can constrain the fraction of short-timescale events that would be due to free-floating 
planets given our assumed planet model.

\begin{figure}
	\epsscale{1.15}
	\plotone{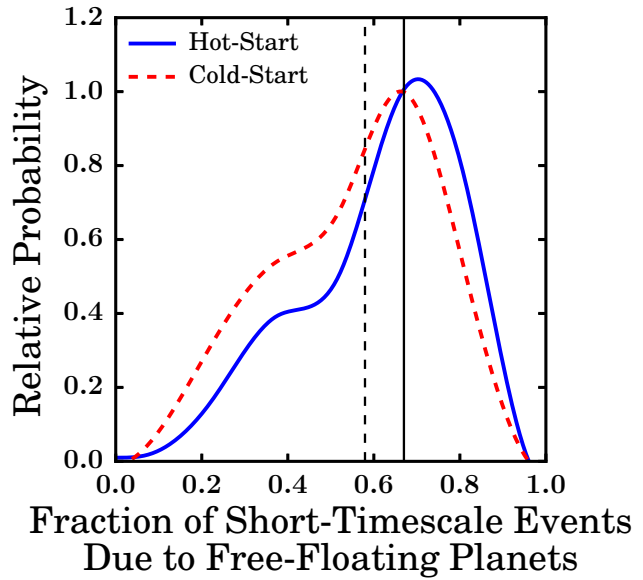}
	\caption{The fraction of short-timescale ($t_E<2~$days) microlensing events 
	that must be due to free-floating planets, $f_\mathrm{ff}$, 
	              for our analyses that assume either 
	               ``hot-start'' \citep[blue;][]{Baraffe2003} or ``cold-start'' 
	               \citep[red;][]{Fortney2008} planet evolutionary models. The vertical, 
	               black lines mark the median values of these posterior distributions.
		\label{fig:frac_ste_ffp}}
\end{figure}

For each planet population we fit to the observed timescale distribution, we determine the 
number of residual events with $0.3< t_E / \mathrm{days} < 2$ and divide by the 
number of observed events in this same range of $t_E$ to compute the fraction of such 
events which are expected to be due to free-floating planets, $f_\mathrm{ff}$. 
We plot the posterior distribution of $f_\mathrm{ff}$ in Figure~\ref{fig:frac_ste_ffp} and 
report the corresponding median values, 68\%, and 95\% confidence intervals in 
Table~\ref{tab:tab2}. The 
posterior for the ``cold-start'' case is shifted slightly towards lower $f_\mathrm{ff}$, as expected, 
but it is not significantly different from the ``hot-start'' case.

\begin{table*}
	\caption{\label{tab:tab2} Median values, 68\%, and 95\% confidence intervals on 
	both the fraction of short-timescale events due to free-floating planets, $f_\mathrm{ff}$, 
	and the number of free-floating planets relative to main-sequence stars, $N_\mathrm{ff}$. 
	We report these values for our analyses that assume either ``hot-start'' 
	\citep{Baraffe2003} or ``cold-start'' \citep{Fortney2008} planet evolutionary models.}
	\centering
	\begin{tabular}{c||c|c|c|c}
		\hline \hline
		& Planet Evolutionary  &  Median & 68\% Confidence & 95\% Confidence\\
		& Model                     &  Value    & Interval              & Interval  \\
		\hline
		\multirow{2}{*}{$f_\mathrm{ff}$} & ``Hot-Start'' & $0.67$ & $0.44-0.78$ & $0.23-0.85$ \\
		\cline{2-5}
	                                                       & ``Cold-Start'' & $0.58$ & $0.40-0.74$ & $0.14-0.83$ \\
	     \hline
	     \multirow{2}{*}{$N_\mathrm{ff}$} & ``Hot-Start'' & $1.4$ & $0.94-1.7$ & $0.48-1.8$ \\
		\cline{2-5}
	                                                       & ``Cold-Start'' & $1.2$ & $0.86-1.6$ & $0.29-1.8$ \\
		\hline\hline
	\end{tabular}
\end{table*}

\begin{figure}
	\epsscale{1.15}
	\plotone{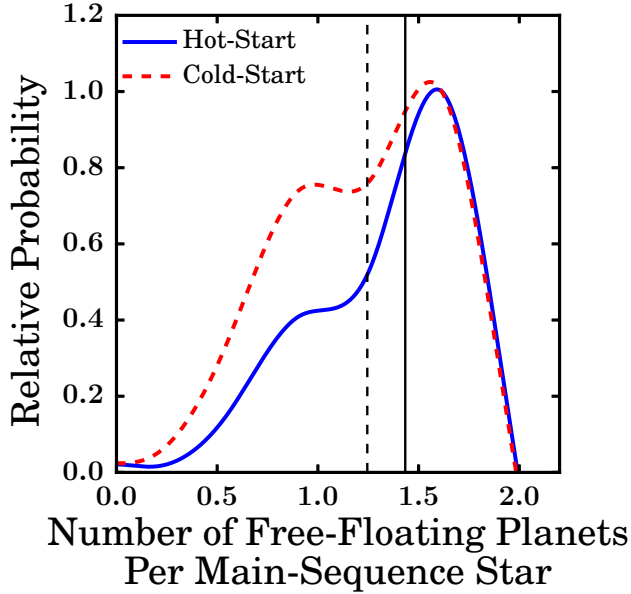}
	\caption{The number of free-floating planets per main-sequence star, 
	$N_\mathrm{ff}$, required 
	to explain the residual short-timescale ($t_E<2~$days) microlensing events 
	after fits of our canonical LMF and populations of wide-separation, bound planets 
	are subtracted for our analyses that assume either 
	               ``hot-start'' \citep[blue;][]{Baraffe2003} or ``cold-start'' 
	               \citep[red;][]{Fortney2008} planet evolutionary models. The vertical, 
	               black lines mark the median values of these posterior distributions. 
	               Estimating this quantity requires an assumption about the mass 
	               function of free-floating planets. Here, we have chosen a delta function at 
	               a mass of $2~M_\mathrm{Jup}$ (see text for discussion).
		\label{fig:num_ste_ffp}}
\end{figure}

\begin{figure}[!t]
	\epsscale{1.15}
	\plotone{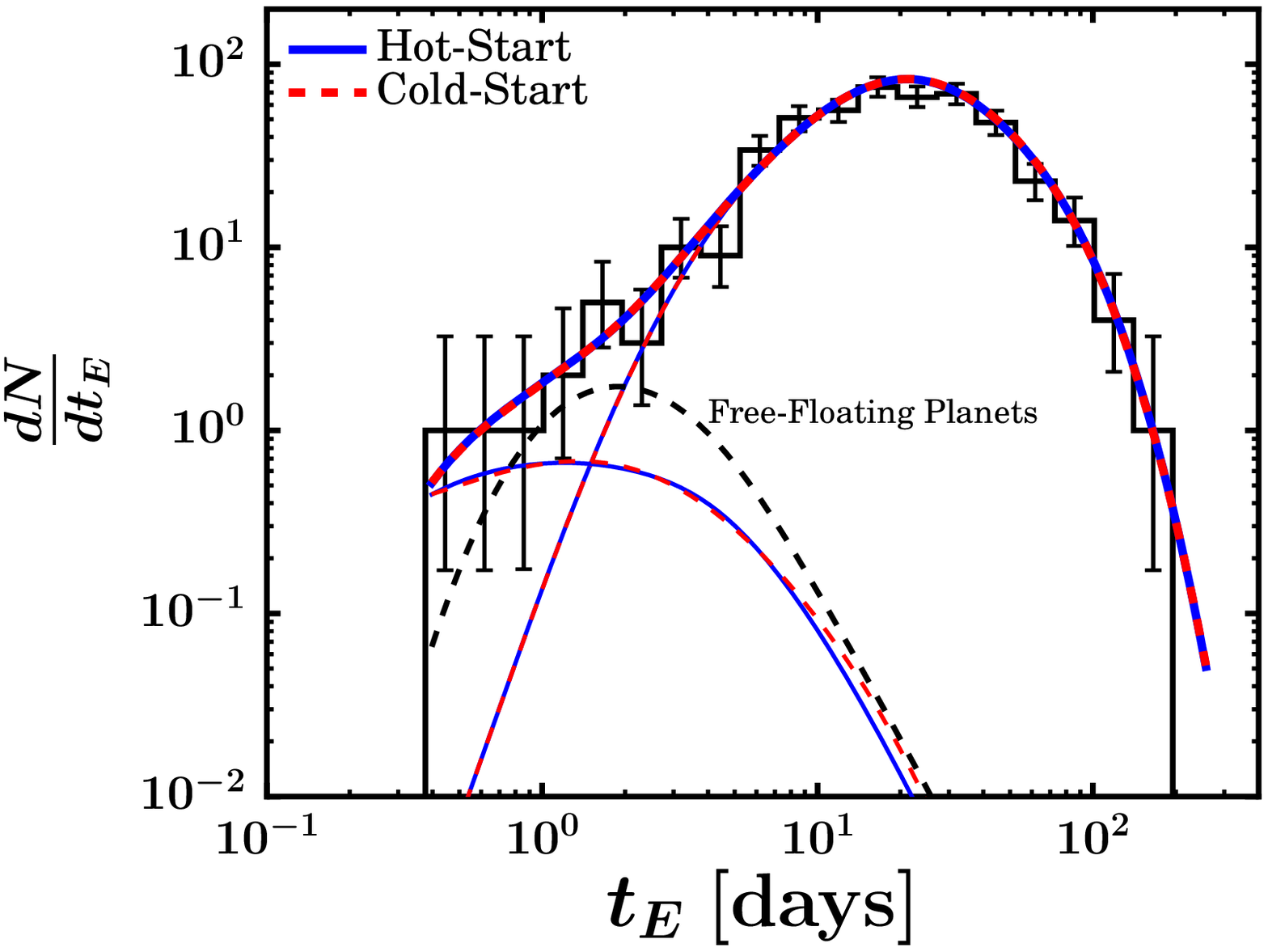}
	\caption{Maximum likelihood fits to the observed timescale distribution 
	              \citep[black histogram;][]{Sumi2011} for our canonical LMF, 
	              a population of bound, 
	              wide-separation planets that is consistent with results from radial 
	              velocity, microlensing, and direct imaging surveys \citep{Clanton2016}, 
	              assuming either  
	               ``hot-start'' \citep[blue lines;][]{Baraffe2003} or ``cold-start'' 
	               \citep[red lines;][]{Fortney2008} planet evolutionary models, and a population of 
	               free-floating planets whose mass function is a $\delta$ function at 
	               $2~M_\mathrm{Jup}$ (black dashed line). The thick 
	               lines show the expected timescale distribution from all lenses, while 
	               the thin lines show the expected contributions from bound planets (the curves 
	               peaking at 
	               shorter timescales) and brown dwarfs, main-sequence stars, and 
	               remnants (the curves peaking at longer timescales). For these  
	               maximum likelihood 
	               fits, wide-separation, bound planets account for roughly 2.9 of the 10 
	               observed 
	               short-timescale ($t_E < 2~$days) events in both the ``hot-start'' and 
	               ``cold-start'' cases, brown dwarfs account for about one event, and free-floating planets 
	               make up the difference.
		\label{fig:max_lhood_te_dist_fit_with_ffp}}
\end{figure}

\begin{figure}
	\epsscale{1.15}
	\plotone{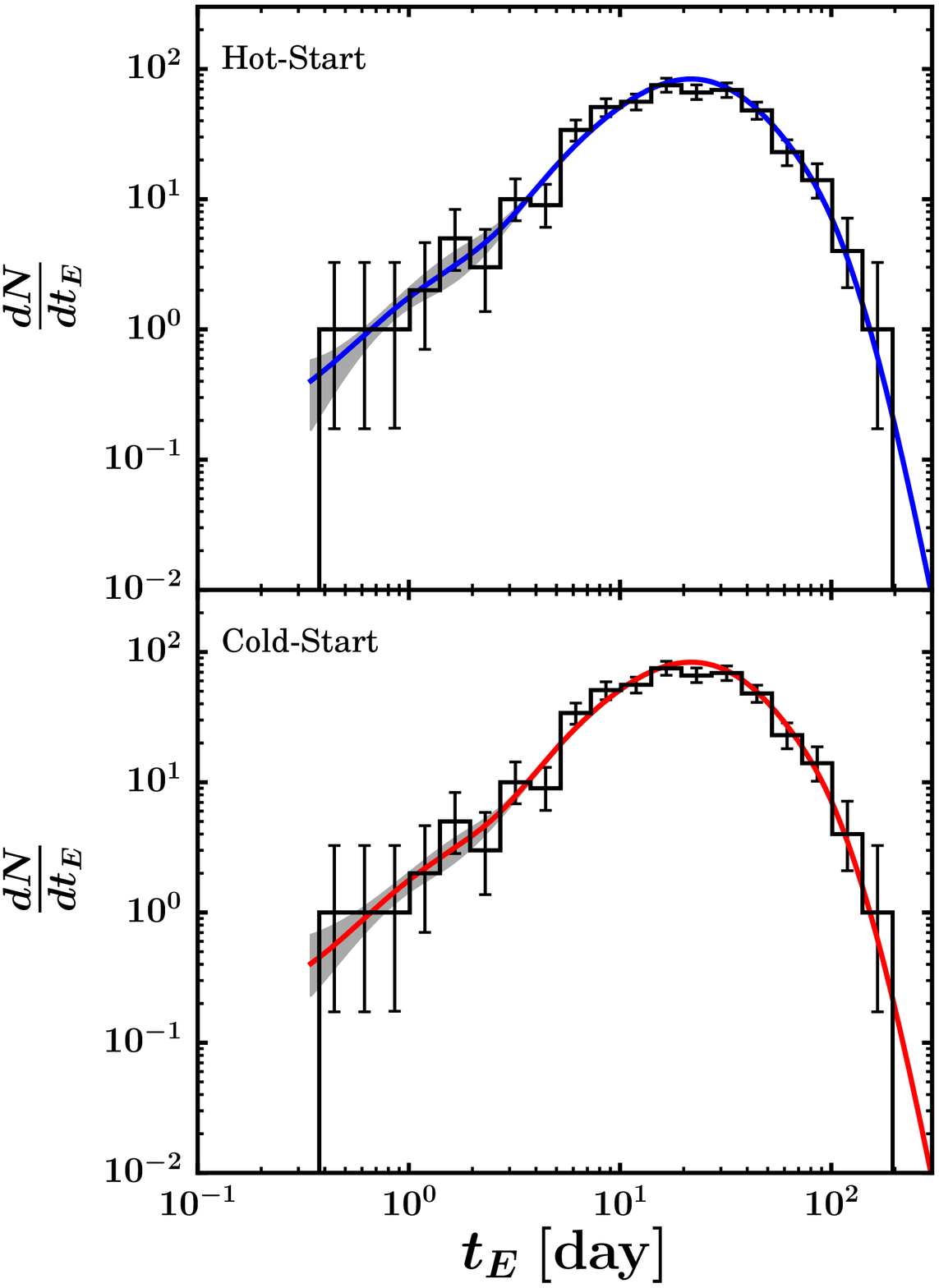}
	\caption{Maximum likelihood and 
	              68\% confidence interval fits to 
	              the observed timescale distribution 
	              \citep[black histograms;][]{Sumi2011} for our canonical LMF, 
	              a population of bound, 
	              wide-separation planets that is consistent with results from radial 
	              velocity, microlensing, and direct imaging surveys \citep{Clanton2016}, 
	              assuming either  
	               ``hot-start'' \citep[top panel;][]{Baraffe2003} or ``cold-start'' 
	               \citep[bottom panel;][]{Fortney2008} planet evolutionary models, and a 
	               population of free-floating planets whose mass function is a $\delta$ 
	               function at $2~M_\mathrm{Jup}$.
		\label{fig:te_dist_fit_with_ffp_range}}
\end{figure}

In order to turn this fraction, $f_\mathrm{ff}$, into an actual number of free-floating planets 
(relative to main-sequence stars, for example), we must assume a form for their mass function. 
To this end, we assume that the free-floating planet mass function is given by a Dirac delta
 function, $\delta (m_\mathrm{p,\; ff} / M_\mathrm{Jup} - 2)$. We chose to center the 
 delta function at $2~M_\mathrm{Jup}$ as such a free-floating planet population lead to a 
 timescale distribution that most closely matches (by eye) the residuals obtained from 
 subtracting off our LMF and the population of wide-separation, bound planets as described 
 in the previous section. Admittedly, this is a rough calculation, however given the level of 
 precision of 
 this study, we do not believe a more careful analysis is currently warranted (especially given 
 the fact that we currently have no constraints on the actual form of the free-floating planet mass 
 function that we must adopt). The resultant posterior on the number of free-floating planets per 
 main-sequence star is plotted in Figure~\ref{fig:num_ste_ffp} and the corresponding median 
 values, 68\%, and 95\% confidence intervals are reported in 
 Table~\ref{tab:tab2}. We plot the maximum likelihood fits for the ``hot-'' 
 and ``cold-start'' analyses, including the contribution from free-floating planets, 
 under this assumption 
 of a delta-function mass distribution at $2~M_\mathrm{Jup}$ in 
 Figure~\ref{fig:max_lhood_te_dist_fit_with_ffp}, and we show the range of fits 
 in the 68\% confidence 
 interval in Figure~\ref{fig:te_dist_fit_with_ffp_range}.
 
 The median number of free-floating planets per main-sequence star we find, 
$N_\mathrm{ff}=1.4^{+0.30}_{-0.46}$ ($N_\mathrm{ff}=1.2^{+0.40}_{-0.34}$) 
for the ``hot-start'' (``cold-start'') case, 
is quite a large number that seems difficult to explain with any known formation 
mechanism (see Section~\ref{sec:sec1} for discussion on the formation channels for 
free-floating planets). However, more ``comfortable'' values of 
$N_\mathrm{ff}=0.48$ ($N_\mathrm{ff}=0.29$; ``cold-start'') are allowed to within 95\% confidence 
and could perhaps be easier to explain. Furthermore, these values are sensitive to 
a number of assumptions, most notably the free-floating planet mass function and 
the model for the population of wide-separation, bound planets. The remainder of this 
section is devoted to discussion of these two primary sources of uncertainty.

\textit{Free-Floating Planet Mass Function:} Without direct lens mass measurements 
for each of the short-timescale events, the only constraining power currently 
available on the free-floating planet mass function is, in fact, contained in the 
observed microlensing event timescale distribution. In order to constrain the 
free-floating planet mass function using the timescale distribution, prior knowledge 
of which of the short-timescale events are actually due to free-floating planets 
would be required. With such knowledge, one could perform a model comparison 
of fits to the observed timescale distribution assuming different forms for the 
free-floating planet mass function. Unfortunately, we do not know exactly which 
events are caused by truly unbound planets (due to fundamental degeneracies 
that affect a majority of microlensing observations; see 
\citealt{Gaudi2012} and references therein) and the number of short-timescale 
$t_E\lesssim 2~$days is small, making such a study difficult. This paper is an 
attempt to address the first of these issues by simulating microlensing events of 
wide-separation, bound planets to determine (statistically) the fraction of the 
short-timescale events that are caused by free-floating planets. Of course, the 
results presented herein are therefore dependent on our assumed model 
of bound planets.

Data from ongoing and future microlensing surveys will allow direct measurements 
of both the frequency and mass function of free-floating planets, as well as their 
spatial distribution within our Galaxy. The recent \textit{K2} Campaign 9 (\textit{K2}C9)
consisted of a survey toward the Galactic bulge 
\citep{Henderson2015}. For 
short-timescale microlensing events observed simultaneously from \textit{Kepler} 
and ground-based observatories (such that we see two distinct source trajectories), 
it is possible (for some events) to directly measure the lens mass 
and distance and obtain better constraints on the existence of a primary (i.e. host 
star) since \textit{Kepler} provides precise, continuous observations 
\citep[see][]{Henderson2015,Henderson2016}. However, given the short, 
$\sim 80~$day duration of \textit{K2}C9, the sample size will likely be too small to make 
population-level inferences about free-floating planets other than (at least limits) on their 
occurrence rates (recall that the event rate scales as $\Gamma \propto M_L^{1/2}$). 
Indeed, \citet{Penny2016} predict that \textit{K2}C9 will detect between 1.4 and 7.9 
microlensing events due to free-floating planets (assuming 1.9 free-floating planets per 
main-sequence star per the \citealt{Sumi2011} result). Of these expected detections, 
\citet{Penny2016} predict that for between 0.42 and 0.98 it will be possible to gain a 
complete solution (i.e. to measure both finite-source effects and microlens parallax). 
Given the results we present in this paper, these numbers would be smaller by a factor 
of $\sim0.6$ (refer to Table~\ref{tab:tab2}), and thus it is unlikely \textit{K2}C9 will 
actually directly measure the lens mass in a short-timescale event.

Fortunately, the microlensing survey of the \textit{Wide-Field InfraRed Survey 
Telescope} 
\citep[hereafter \textit{WFIRST};][]{Spergel2015} will ultimately, when combined with 
ground-based observations, provide the necessary data to directly 
measure frequencies, masses, 
and distances for a large sample of free-floating planets with masses down to that 
of Mars (see \citealt{Gould2003} and \citealt{Yee2013}, who demonstrate  
that simultaneous observations from the ground and \textit{WFIRST} at L2 will enable 
the measurement of the parallax of planetary events). 
Depending on the exact occurrence rates, \textit{WFIRST} will detect 
$\sim$hundreds to $\sim$thousands of free-floating planets 
\citep[see Table~2-6 of ][]{Spergel2015}.

\textit{The Population of Wide-Separation, Bound Planets:} The model we assume 
in this paper is a joint power-law distribution function in planet mass and semimajor 
axis that \citet{Clanton2016} demonstrate to be consistent with results from radial 
velocity, microlensing, and direct imaging surveys (the caveats and uncertainties of 
which are laid bare in Section~6 of \citealt{Clanton2016}). However, the region of planet 
parameter space we examine in this paper ($m_p\lesssim M_\mathrm{Jup}$; 
$a\gtrsim 10~$AU) is not directly constrained by any observations. We have 
implicitly assumed that our distribution function extrapolates into this region of 
parameter space. It could be the case that the form of the planet mass function 
depends on semimajor axes for $a\gtrsim 10~$AU, which could significantly alter  
our conclusions. For example, if no planets with masses 
$m_p\gtrsim M_\mathrm{Jup}$ form beyond $\sim 10~$AU but there is an abundance 
of slightly less massive planets, we could easily explain most, if not all, the 
short-timescale microlensing events with bound planets and still satisfy results from all 
radial velocity, microlensing, and direct imaging surveys of M stars.

Future observations will provide the necessary sensitivity to test the planetary mass 
function at wide-separations and determine whether or not the 
mass function measured by microlensing surveys extends further out (as we have assumed 
to be the case in this paper). The 
\textit{James Webb Space Telescope} \citep[\textit{JWST};][]{Gardner2006} is expected 
to have the 
capability to achieve contrasts of $\sim 10^{-5}$ at angular separations 
$\gtrsim 0.6~$arcseconds for 
observations at $\sim4.5~\mu$m with NIRCam \citep[and even greater sensitivity at 
larger separations;][]{Horner2004,Krist2007}. A survey of nearby, young M stars 
with \textit{JWST}/NIRCam as proposed by \citet{Schlieder2016} has the potential to 
probe down to masses of $\sim 0.1~M_\mathrm{Jup}$ at separations of $\sim 10~$AU, 
complementary (and perhaps with some overlap) to microlensing surveys.

\section{Summary}
\label{sec:sec6}
In this paper, we attempt to explain the observed overabundance of short-timescale 
($t_E<2~$days) microlensing events with populations of wide-separation, bound 
exoplanets that are known to be simultaneously consistent with results from radial 
velocity, microlensing, and direct imaging surveys. We select planetary 
systems from such populations that we (statistically) expect not to show evidence of 
a primary (i.e. host star) in their microlensing light curves, either via low-magnification 
bumps or anomalies near the peak of the light curve due to close approaches to, or 
crossings of, the planetary caustics. We fit the observed timescale distribution reported 
by \citet{Sumi2011} with these planetary systems and a primary lens mass function 
consisting of brown dwarfs, main-sequence stars, white dwarfs, neutron stars, and black 
holes. We find that wide-separation, bound planets can explain some of the short-timescale 
events, but (assuming our joint power-law planet distribution function in mass and 
semimajor axis presented in \citealt{Clanton2016} is correct) free-floating planets must 
account for a fraction of the short-timescale events of either 
$f_\mathrm{ff}=0.67$ (0.23--0.85 at 95\% confidence) for ``hot-start'' planet evolutionary 
models \citep{Baraffe2003} or 
$f_\mathrm{ff}=0.58$ (0.14--0.83 at 95\% confidence) for ``cold-start'' models 
\citep{Fortney2008}.

The fraction of short-timescale events due to free-floating planets is 
the most robust statistic we can infer from the available data (see Section~\ref{sec:sec5}). 
In order to determine 
an occurrence rate of free-floating planets, we must necessarily assume something 
about their mass function (for which there is currently no observational constraints). 
We choose to adopt a free-floating planet mass function that is a delta function at 
$2~M_\mathrm{Jup}$, as this (roughly) reproduces the residual timescale distribution 
after subtraction of our canonical LMF and wide-separation, bound planets that are not 
expected to show evidence of a primary. Under this assumption, we compute the number 
of free-floating planets per main-sequence star and find a median value 
$N_\mathrm{ff}=1.4$ (0.48--1.8 at 95\% confidence) 
in the ``hot-start'' case and 
$N_\mathrm{ff}=1.2$ (0.29--1.8 at 95\% confidence) 
for the ``cold-start'' case.

These values are slightly lower than that suggested by \citet{Sumi2011} of 
$1.8^{+1.7}_{-0.8}$, but still seem difficult to explain given our current understanding of 
formation channels for free-floating planets. Our results also suggest occurrence rates of 
free-floating planets that is higher by a large factor than 
that inferred by the SONYC imaging 
survey of NGC 1333 \citep{Scholz2012b}, but quite a bit lower than that inferred by the 
photometric survey of the $\rho$ Oph cloud core by \citet{Marsh2010}. Potential reasons 
for the differences in frequencies of free-floating planets between our results and those of 
imaging surveys are 1) imaging surveys are probing a different population of free-floating, 
planetary-mass objects, 2) these frequencies are heavily 
dependent on the local environmental conditions, 3) the imaging surveys, which 
are only typically sensitive to 
objects more massive than about a couple Jupiter masses, lack the sensitivity to probe the 
free-floating planet population inferred by microlensing.

Future observations will be critical to further elucidate the true abundance and 
demographics of free-floating 
planets. Direct mass measurements of a statistically-significant sample of short-timescale 
microlensing events will allow us to infer 
the mass function of free-floating planets. It is unlikely that there will be any lens mass 
measurements from short-timescale microlensing events from the \textit{K2} Campaign 9 
dataset (see Section~\ref{sec:sec5}), 
but as \citet{Penny2016} point out, \textit{K2}C9 can still test the hypothesis that these events 
are, in fact, due to planetary-mass objects, and if so, whether or not they are bound to stars. Ultimately the microlensing survey of \textit{WFIRST} will provide robust measurements 
of the free-floating planet mass function, their occurrence rates, and their Galactic 
distribution \citep{Spergel2015}.

\acknowledgments
We thank Takahiro Sumi and Radek Poleski for helpful conversations. 
This research has made use of NASA's Astrophysics Data System and was 
partially supported by NSF CAREER Grant AST-1056524. Work by CDC was 
supported in part by an appointment to the NASA Postdoctoral Program at 
Ames Research Center that is administered by the 
Universities Space Research Association through a contract with NASA.

\end{document}